\documentclass[10pt,aps,pra,twocolumn,superscriptaddress]{revtex4-2}

\usepackage[pages=all, color=black, position={current page.south}, placement=bottom, scale=1, opacity=1, vshift=5mm]{background}
\SetBgContents{
	\tt 
}      

\usepackage[margin=1in]{geometry} 

\usepackage{amsmath}
\usepackage{amsthm}
\usepackage{amssymb}
\usepackage{times}
\usepackage{braket}
\usepackage{ulem}

\usepackage[utf8]{inputenc}
\usepackage{svg}
\usepackage{hyperref}

\usepackage[english]{babel}

\usepackage{graphicx, color}
\graphicspath{{fig/}}

\usepackage{algorithm, algpseudocode} 
\usepackage{mathrsfs} 
\usepackage{lipsum}

\begin{document}

\title{Photonic fusion operations transform partial distinguishability}

\author{S. N. van den Hoven}
\affiliation{MESA+ Institute for Nanotechnology, University of Twente, P.~O.~box 217, 7500 AE Enschede, The Netherlands} 

\author{G.B. Lamers}
\affiliation{Department of Applied Physics and Science Education, Eindhoven University of Technology, P.O. Box 513, 5600 MB
Eindhoven, The Netherlands}

\author{J. J. Renema}
\affiliation{MESA+ Institute for Nanotechnology, University of Twente, P.~O.~box 217, 7500 AE Enschede, The Netherlands} 
\affiliation{Department of Applied Physics and Science Education, Eindhoven University of Technology, P.O. Box 513, 5600 MB
Eindhoven, The Netherlands}
\affiliation{Department of Electrical Engineering, Eindhoven University of Technology, P.O. Box 513, 5600 MB Eindhoven, The Netherlands}

\date{
	\today
}

\begin{abstract}
    Photonic fusion operations are a central primitive in photonic quantum information processing and are commonly characterized by their success probability and resource cost. Here, we identify an additional way in which different implementations of fusion operations differ, namely in how they transform partial distinguishability. We consider distinguishability originating both from imperfect entangled resource states and from imperfect single photons. We find that different successful fusion events can induce markedly different distinguishability transformations depending on the fusion protocol, heralding outcome, and error model. Under a collective resource-state error model, we identify a modification to an existing fusion protocol that improves the fidelity of the remaining state on average. Generally, however, these transformations adversely affect distinguishability, and many successful fusion outcomes are accompanied by distinguishability degradation. Our results show that partial distinguishability is dynamically shaped by interference and measurements and should be considered alongside overall fusion success probability when evaluating photonic fusion protocols.

\end{abstract}

\maketitle
\section{Introduction}

Photonic quantum technologies constitute one of the most promising approaches towards large-scale quantum information processing due to their low decoherence rates, integrability with the telecom industry, and natural suitability for transmitting quantum information over long distances. Recent years have witnessed substantial experimental progress, including advances in deterministic single-photon sources, integrated photonic platforms, and the generation of increasingly complex photonic resource states \cite{Thomas2022, Albrechtsen2026, Alexander2025, Meng2025}. As photonic systems continue to scale in size and complexity, understanding the effect of realistic imperfections on multiphoton quantum states becomes increasingly important.

A central feature underlying photonic quantum information processing is multiphoton interference \cite{Fox2006QuantumOptics}. Since photons do not naturally interact, effective nonlinearities must instead arise from interference effects combined with measurements. This mechanism underlies linear optical quantum computing \cite{knill2001scheme}, measurement-based approaches using cluster states \cite{PhysRevLett.86.5188}, fusion-based quantum computation (FBQC) \cite{bartolucci2023fusion, PRXQuantum.4.020303}, and quantum networking architectures based on photonic entanglement distribution \cite{BARRAL2025100747, Saied2024AQN, fxcg-xxry}. 

Among the most important interference-based primitives are Bell measurements, commonly referred to as \textit{fusions} \cite{PhysRevLett.95.010501}. Fusions can be used to construct larger entangled resource states from smaller building blocks, to perform entanglement-swapping operations between distant nodes, and, in fusion-based architectures, to drive the computation itself. They therefore play a central role in a wide range of photonic quantum-information protocols. In particular, large-scale FBQC architectures rely on vast numbers of fusion operations both during resource-state preparation and throughout the computation itself. Since linear-optical Bell measurements are inherently probabilistic, substantial effort has been devoted to increasing their success probability through the use of ancillary resource states \cite{PhysRevLett.113.140403,PhysRevA.84.042331}. Consequently, fusion protocols are commonly characterized by their success probability and ancillary resource requirements.

Because fusion operations rely directly on multiphoton interference, their performance is sensitive to deviations from ideal bosonic behaviour. Such deviations, commonly referred to as \textit{partial distinguishability}, arise when photons carry residual which-way information through internal degrees of freedom (e.g. spectrum or polarization). Partial distinguishability degrades quantum interference \cite{hong1987measurement, PhysRevLett.118.153603Triadphase, Seron2023, g28d-jzgj} and can undermine photonic quantum information processing by enabling efficient classical simulation of boson sampling \cite{renema2018efficient, PhysRevA.111.052448} or by introducing errors during entangling operations \cite{sparrow2018phd_thesis,Saied2024AQN,PhysRevA.73.062312}.

Recent work has shown that carefully designed interference experiments and partial measurements can probabilistically improve the quality of single photons through \textit{distillation}-like processes \cite{marshall2022distillation, sparrow2018phd_thesis, saied2024general, somhorst2024photon, faurby2024purifying, somhorst2026belowthresholderrorreductionsingle, hoch2025optimaldistillationphotonicindistinguishability}. These results demonstrate that measurements do not only reveal imperfections but can also actively transform them. This naturally raises the question of whether fusion operations themselves possess analogous distinguishability-transformation properties.

In this work, we investigate how photonic fusion protocols transform partial distinguishability. Rather than considering distinguishability solely as a source of reduced interference visibility, we study its evolution conditioned on successful fusion events and ask whether fusion operations themselves can exhibit distillation-like behaviour. Specifically, we investigate whether different fusion protocols preferentially suppress or amplify distinguishability properties in the surviving state.

To address this question, we study two complementary scenarios. First, we introduce a collective distinguishability model for Bell states and investigate how the fidelity of Bell states evolves when noisy Bell pairs are fused using different protocols with imperfect ancillary resources. Second, we consider a more granular description based on distinguishable single photons and investigate the average effect of fusion operations acting on larger resource states constructed from such photons. Together, these approaches allow us to investigate distinguishability both at the level of entangled resource states and at the level of their underlying constituents.

We find that different fusion protocols can exhibit substantially different behaviour with respect to the distinguishability content of the heralded state. Differences appear not only between different fusion protocols, but also within the same fusion protocol when different heralding patterns, or different error models are considered. Consequently, protocols that appear comparable in terms of success probability and ancillary resource cost can nevertheless differ in how they transform imperfections. Our results therefore identify an additional design consideration for photonic fusion operations that should be considered alongside the traditional metrics of resource overhead and success probability. More broadly, our work highlights the importance of understanding how realistic imperfections evolve through repeated interference processes in large-scale photonic quantum systems.

\section{Preliminaries}

\subsection{Photonic fusions}
In linear optical quantum information processing, destructive and probabilistic Bell measurements, commonly referred to as \textit{fusions} \cite{PhysRevLett.95.010501}, constitute one of the central primitives. Depending on the specific application, fusion operations can be used to construct large entangled resource states from smaller building blocks \cite{bartolucci2021creation, Lee2023graphtheoretical, 8l5g-x3b7, bartolucci2025comparisonschemeshighlyloss}, to perform entangling gates within a computation \cite{bartolucci2023fusion, PRXQuantum.4.020303, PhysRevLett.131.120603}, or to distribute entanglement across distant nodes \cite{Wehner2018, BARRAL2025100747, Saied2024AQN}.

A central challenge for photonic architectures is that linear-optical Bell measurements are fundamentally probabilistic. The maximum success probability of standard linear-optical Bell measurements with ancillary modes prepared in the vacuum state is limited to $50\%$ \cite{Calsamiglia2001}. The use of additional ancillary resource states can increase this success probability \cite{PhysRevLett.113.140403,PhysRevA.84.042331}. Consequently, different fusion protocols are commonly characterized in terms of their ancillary resource requirements and achievable success probabilities. However, these metrics alone do not completely characterize the performance of a fusion operation in the presence of realistic imperfections.

Loss errors affect fusion operations in a comparatively direct manner. Successful Bell measurements rely on observing specific multi-photon detection signatures, and the loss of a single photon can alter these signatures sufficiently to convert a successful event into either a failure event or an incorrect outcome. Consequently, fusion protocols employing larger ancillary resource states generally become increasingly sensitive to loss, since the probability that all required photons survive decreases with the total photon number.

The effect of partial distinguishability is more subtle. Several works have shown that distinguishability effects can often be interpreted as inducing effective measurement errors \cite{saied2024general,sparrow2018phd_thesis,Saied2024AQN,PhysRevA.73.062312}. Such descriptions provide an intuitive and practically useful way of understanding the effect of distinguishability on quantities directly related to measurement outcomes. However, they do not generally capture the effect that a fusion operation may have on the distinguishability properties of the remaining heralded state. Consequently, while distinguishability-induced measurement errors have received considerable attention, comparatively little is known about how different fusion operations transform the distinguishability content of the surviving state itself.

\subsection{Partial distinguishability}
\label{subsec: Partial dist}

Physically, partial distinguishability can be understood as a mismatch in one or more internal photonic degrees of freedom such as temporal profile, spectrum, spatial mode, or polarization. Consequently, photons that are intended to interfere are no longer perfectly identical. The mathematical framework describing distinguishability effects in many-particle interference has been studied extensively \cite{tichy2015sampling, shchesnovich2015partial}. This framework has been used to study the effect of partial distinguishability on measurement outcomes \cite{saied2024general,sparrow2018phd_thesis,Saied2024AQN,PhysRevA.73.062312}. However, such descriptions do not capture the effect that a measurement itself may have on the remaining heralded state. 

To isolate this effect, we employ simplified distinguishability models that retain the essential physics while remaining analytically tractable. Throughout this work we use orthogonal distinguishability models, in which imperfect photons occupy internal modes orthogonal to the desired target mode. While such models neglect the continuous structure of realistic mode mismatch, they have been widely adopted as a phenomenological model of distinguishability and have proven useful for describing and simulating experimentally relevant photonic systems\cite{sparrow2018phd_thesis, saied2024general, Saied2024AQN, marshall2022distillation, somhorst2024photon, somhorst2026belowthresholderrorreductionsingle, schadow2026certificationlinearopticalquantum, moylett2019classically, annoni2025incoherentbehaviorpartiallydistinguishable, renema2018efficient}.

We first employ the orthogonal bad bit (OBB) model introduced in Ref.\cite{sparrow2018phd_thesis}. In this model, a single-photon state is represented by a mixture consisting of two components. With probability $(1-\epsilon)$ the photon occupies the desired internal mode $\psi_0$, while with probability $\epsilon$ it occupies an orthogonal undesired mode $\psi_i$,

\begin{equation}
\rho_{\mathrm{OBB}}
=
(1-\epsilon)
|1_{\psi_0}\rangle\langle1_{\psi_0}|
+
\epsilon
|1_{\psi_i}\rangle\langle1_{\psi_i}|.
\end{equation}

Following the standard OBB prescription, all undesired modes are assumed mutually orthogonal,

\begin{equation}
\langle \psi_i | \psi_j \rangle
=
\delta_{ij},
\end{equation}

such that every faulty photon is distinguishable both from photons in the target mode and from all other faulty photons.

In addition, to investigate distinguishability effects directly at the level of entangled resource states, we introduce a collective extension of this model, which we refer to as the orthogonal bad Bell (OBBell) model. Rather than assigning distinguishability independently to individual photons, distinguishability is assigned collectively to the Bell pair itself. The state is therefore described by

\begin{equation}
\rho_{\mathrm{OBBell}}
=
(1-\epsilon)
|\Phi^{+}_{\psi_0}\rangle
\langle\Phi^{+}_{\psi_0}|
+
\epsilon
|\Phi^{+}_{\psi_i}\rangle
\langle\Phi^{+}_{\psi_i}|,
\end{equation}

where

\begin{equation}
|\Phi^{+}_{\psi_i}\rangle
=
\frac{
a^{\dagger}_{1,\psi_i}
a^{\dagger}_{3,\psi_i}
+
a^{\dagger}_{2,\psi_i}
a^{\dagger}_{4,\psi_i}
}{
\sqrt{2}
}
|0\rangle 
\end{equation}
denotes a Bell state of two dual rail qubits defined over modes $\{1,2\}$ and $\{3,4\}$ respectively.

Importantly, the photons belonging to a faulty Bell state occupy the same internal mode and therefore remain perfectly indistinguishable with respect to one another. However, photons belonging to different faulty Bell states are assumed to occupy mutually orthogonal modes.

The OBBell model can be viewed as a collective generalization of the OBB model and captures situations in which imperfections originate from the resource-state generation process itself and therefore affect all photons within the resource state simultaneously. This may arise, for example, in deterministic generation schemes based on quantum emitters, where fluctuations during a single emission event introduce a common mode mismatch shared by all emitted photons afterwards \cite{Daggett2026}.

More generally, the OBBell model represents a specific instance of a broader class of collective distinguishability models. For an arbitrary entangled photonic resource state $|\chi\rangle$, one may define a corresponding orthogonal bad resource-state model according to

\[
\rho_{\mathrm{OB}\chi}
=
(1-\epsilon)
|\chi_{\psi_0}\rangle\langle\chi_{\psi_0}|
+
\epsilon
|\chi_{\psi_i}\rangle\langle\chi_{\psi_i}|,
\]

where all photons belonging to the faulty component occupy a common internal mode $\psi_i$, while faulty states originating from different preparation events are assumed mutually orthogonal.

Throughout this work we primarily consider two specific realizations: the Bell-state model OBBell and its 4-GHZ counterpart OBGHZ.

\section{Distillation properties of fusion protocols}
\label{sec: distillation properties}

We now investigate how different fusion implementations transform distinguishability errors in the heralded state. Rather than considering fusion performance solely in terms of success probability and ancillary resource requirements, we study whether successful fusion events can alter the fidelity of the heralded state with respect to the fidelity of the input state. We will investigate both the average evolution of distinguishability as well as the behaviour of individual heralding outcomes.

Throughout this section, we use the term \textit{indistinguishability distillation} to denote situations in which the fidelity of the heralded state exceeds that of the input state, whereas \textit{anti-distillation} refers to a reduction in fidelity following successful fusion.

\subsection{Fusion protocols considered}
\label{subsec: fusion protocols considered}

The complexity of simulating many-photon interference experiments increases rapidly with photon number and distinguishability structure. We therefore restrict our analysis to several representative fusion protocols commonly considered in photonic quantum information processing.

Specifically, we consider

\begin{enumerate}
\item the standard unboosted fusion protocol \cite{PhysRevLett.95.010501},
\item the Grice protocol employing a single Bell state ancilla \cite{PhysRevA.84.042331},
\item the Ewert $\&$ van Loock protocol employing four ancillary single photons \cite{PhysRevLett.113.140403},
\item the Grice protocol employing both a Bell-state and a 4-GHZ ancilla \cite{PhysRevA.84.042331}.
\end{enumerate}

Interestingly, the Ewert $\&$ van Loock protocol can be viewed as belonging to the same family of interferometric constructions as the Grice scheme. The primary distinction lies in the ancillary resource state, which in this case corresponds to four photons transformed according to $\mathrm{H}_4$. To our knowledge this connection has not been explicitly highlighted previously.

The ancillary resource state thus represents one of the principal design freedoms available in boosted fusion protocols. We therefore performed a numerical investigation of alternative ancillary-state configurations while keeping the interferometric structure fixed. While many such configurations realize valid Bell-measurement schemes, their success probabilities and the way they transform distinguishability vary considerably.

For example, one configuration employing $|\Phi^+\rangle^{\otimes 3}$ ancilla states achieves a Bell-measurement success probability of $78.5\%$ in the noiseless limit while requiring only Bell-state resources rather than a 4-GHZ state. A detailed discussion of these alternative constructions is beyond the scope of the present work and is deferred to the Supplementary Material. Nevertheless, these observations suggest that ancillary-state design itself can significantly influence fusion performance.

In the present work, however, we focus specifically on distinguishability propagation. Guided by the numerical investigation described above, we introduce two modified protocols whose ancillary-state structure exhibit particularly favourable distinguishability behaviour.

The first modification replaces the $|\Phi^+\rangle$ Bell-state ancilla of the original Grice proposal with a $|\Psi^+\rangle$ Bell state. The second modification replaces the $|\Phi^+\rangle$ Bell-state plus 4-GHZ ancilla by three $|\Psi^+\rangle$ Bell states. We will refer to these modifications as the altered Grice protocols.

\subsection{Bell-state fusion under collective distinguishability}
\label{sec: OBBell mode, fusing Bell states}

We first consider a setting that isolates distinguishability effects directly at the level of entangled resource states. We subsequently investigate whether the resulting behaviour persists in settings where resource states are generated from imperfect single photons.

Specifically, we first consider the fusion of two Bell states described by the OBBell model introduced in Sec. \ref{subsec: Partial dist}. Ancillary resource states are assumed to follow the corresponding orthogonal distinguishability models: single-photon ancillas follow the OBB model, while Bell and 4-GHZ resource states follow their collective counterparts OBBell and OBGHZ respectively.

To compare protocols featuring ancillary states of different complexity, we assume that the distinguishability parameter increases with resource-state size according to

\[
\epsilon_{\mathrm{OBB}}
=
\frac12
\epsilon_{\mathrm{OBBell}}
=
\frac{1}{4}
\epsilon_{\mathrm{OBGHZ}}.
\]

This choice reflects the expectation that larger entangled resource states generally require more complex preparation procedures and therefore provide additional opportunities for imperfections to accumulate. Such behaviour is qualitatively consistent with experimental observations of decreasing state fidelity with increasing resource-state size in deterministic photonic state generation \cite{Thomas2022}. We emphasize that this parameter choice serves as a physically motivated illustrative model and that our qualitative conclusions remain unchanged under moderate variations of these parameters.

For each fusion protocol, we evaluate the fidelity of the heralded Bell state conditioned on successful fusion outcomes ($\mathcal{F}'$). We evaluate the fidelity for each individual heralding pattern, since individual heralding patterns may possess substantially different distinguishability properties. We characterize both a weighted average based on the probability of observing respective heralding patterns and the distribution of individual heralding patterns.

\begin{figure}[h!]
    \centering
    \includegraphics[width=1\linewidth]{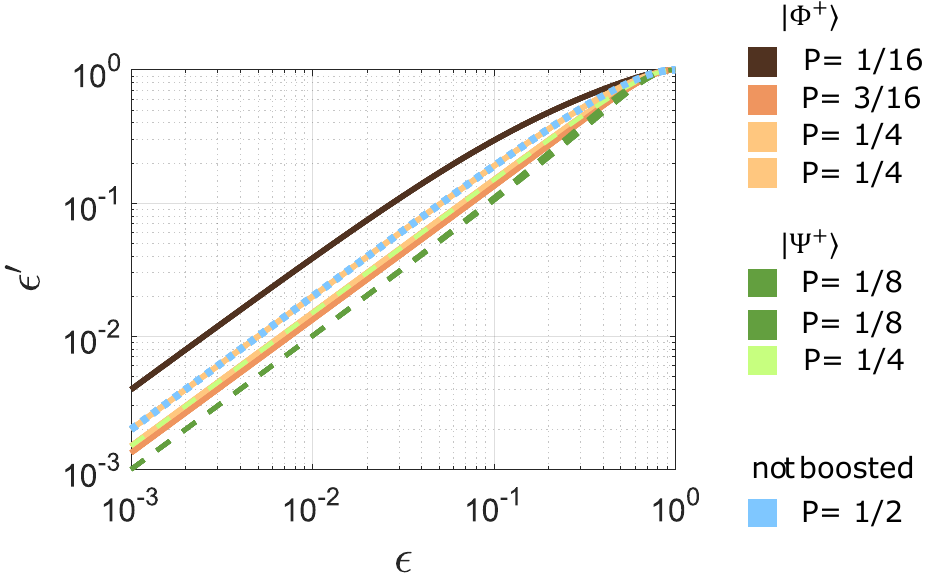}
    \caption{Loglog plot of the infidelity of a Bell state after fusing two Bell states together with various fusion protocols. The output infidelity is plot as a function of the input infidelity of the Bell states. The intersections with the y-axis show the increase in infidelity in the low error regime for various families of heralding patterns. The families of the different heralding patterns for the fusion protocols without ancillary states, with a $|\Phi^+\rangle$ ancilla state and with a $|\Psi^+\rangle$ ancilla state are shown as solid lines with an orange hue, dashed lines with a green hue and a blue dotted line respectively. The different shades of the different lines correspond to the probability of finding a herald pattern from a certain family. The probabilities sum up to $\frac{3}{4}$, $\frac{1}{2}$ and $\frac{1}{2}$ respectively.}
    \label{fig:eps vs eps'}
\end{figure}

Figure (\ref{fig:eps vs eps'}) shows the Bell-state infidelity ($\epsilon'=1-\mathcal{F}'$) as a function of the distinguishability parameter $\epsilon$ for several representative fusion schemes. To preserve visual clarity, we restrict this comparison to the unboosted protocol and Bell-state-boosted protocols employing either $\Phi^+$ or $\Psi^+$ ancillas.

Several observations can be made from Fig.\ref{fig:eps vs eps'}. First, the distinguishability properties of the heralded Bell state differ significantly between fusion protocols despite their similar functionality at the level of Bell-state projections.

Second, distinct herald patterns within a single protocol can exhibit markedly different behaviour. Successful Bell projections therefore do not correspond to a unique distinguishability transformation, but rather to a distribution of distinguishability outcomes.

Third, replacing the ancillary Bell state from $|\Phi^+\rangle$ to $|\Psi^+\rangle$ significantly alters the distinguishability behaviour despite leaving the interferometric structure unchanged. This indicates that distinguishability propagation depends not only on the interferometric structure itself, but also on the detailed choice of ancillary resource states.

Lastly, we note that although the distinguishability propagation differs significantly between protocols and herald patterns, none of the fusion protocols shown in Fig.\ref{fig:eps vs eps'} exhibits net indistinguishability distillation under the OBBell model. 

Taken together, these observations demonstrate that fusion protocols possess distinguishability transformation properties that are largely independent of their Bell-measurement functionality. Consequently, distinguishability propagation should be regarded as an additional design consideration alongside success probability and ancillary resource requirements when comparing fusion protocols.

\begin{figure}[h!]
    \centering
    \includegraphics[width=1\linewidth]{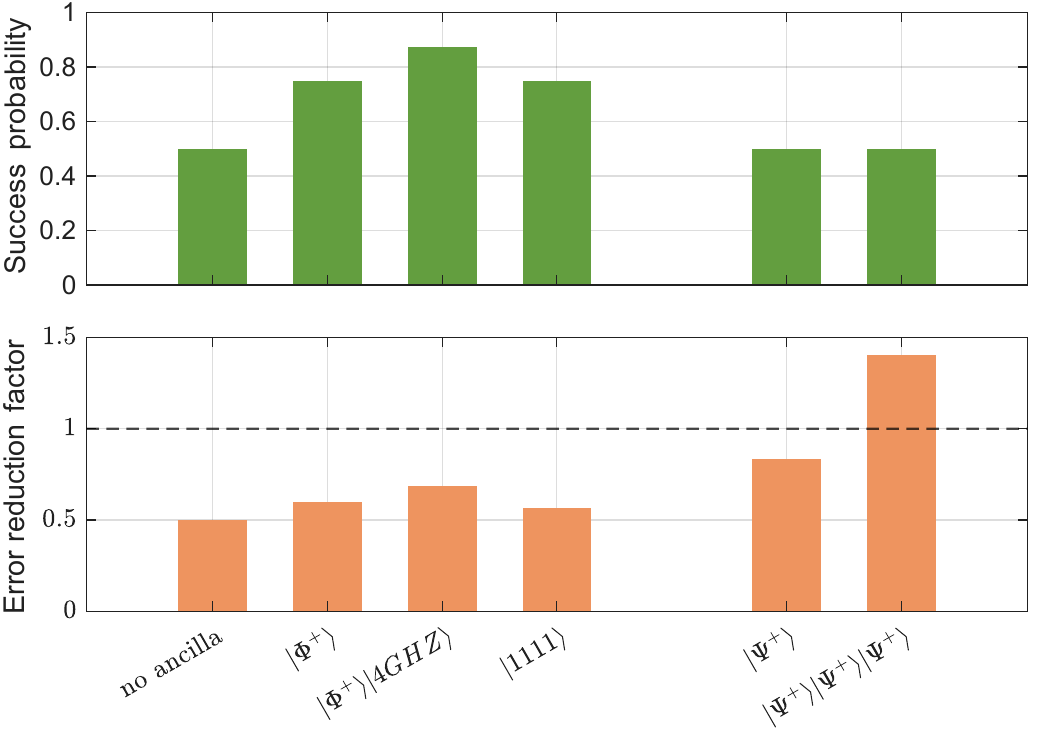}
    \caption{Both the success probability (top) and the expectation value of the error reduction factor (bottom) are shown for various fusion protocols. On the left of the bottom histogram we see  that existing protocols for fusions (up to 6-photon ancillary states) all show error reduction factors below one and hence perform anti-distillation on average, while the altered Grice protocols on the right excel in terms of distillation properties. This comes at the expense of reducing the success probability.}
    \label{fig: histogram}
\end{figure}

To quantify the differences between the distinguishability propagation of different fusion protocols more directly, we now focus on the low-error regime. Figure (\ref{fig: histogram}) summarizes the behaviour of all considered fusion protocols in the limit $\epsilon\rightarrow0$.

For each protocol we show two quantities: 1) the success probability in the absence of distinguishability errors and 2) the average change in Bell-state fidelity induced by successful fusion events. The latter is depicted as an error reduction factor $\dfrac{1-\mathcal{F}}{1-\mathcal{F'}}$.

Notably, for the previously known fusion protocols that use ancillary states with up to $6$ photons, the increase in Bell-measurement success probability obtained through boosting is frequently accompanied by unfavourable indistinguishability distillation properties. While ancillary resource states can be used to increase the probability of obtaining successful Bell projections, they simultaneously introduce additional pathways through which distinguishability can propagate.

Interestingly, modifying the ancillary states can substantially alter this behaviour. In particular, the altered Grice protocols employing $|\Psi^+\rangle$ ancillas exhibit considerably more favourable distinguishability properties than their conventional counterparts. This improvement, however, comes at the cost of a reduced Bell-measurement success probability. These observations indicate that ancillary-state design introduces a trade-off between Bell-measurement success probability and distinguishability propagation.

Figure (\ref{fig: histogram}) shows the expectation value of the indistinguishability distillation properties conditioned on finding a Bell-projection alongside the success probability in the low error limit. Although this is a very useful quantity that can be used to compare various schemes, it does not capture the substantial variation that can occur between individual heralding outcomes.

As discussed above, different herald patterns within the same fusion protocol can exhibit markedly different distinguishability transformations. Consequently, certain successful Bell projections may contribute disproportionately to distinguishability accumulation in larger computations. In some applications, such as resource-state generation, it may therefore be desirable to discard a semi-finished product upon observing herald patterns whose distinguishability properties fall below a specified threshold, even though they correspond to successful Bell measurements in the conventional sense.

\begin{figure}[h!]
    \centering
    \includegraphics[width=1\linewidth,trim=0cm 0cm 0cm 0cm,
    clip]{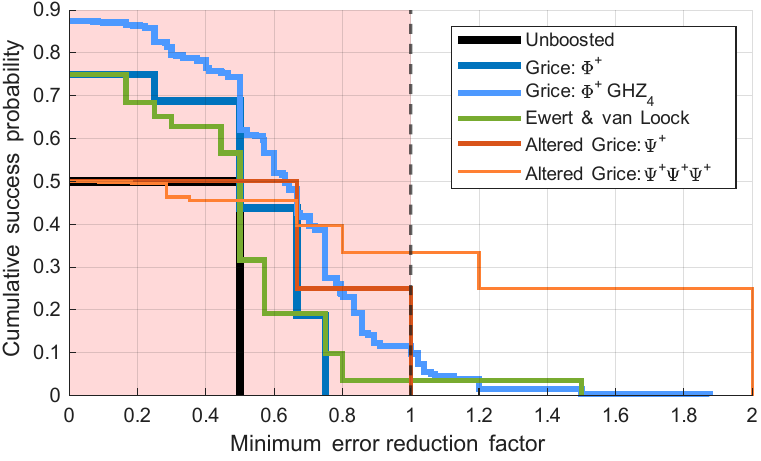}
    \caption{For various fusion protocols, the cumulative success probability is shown as a function of a desired minimum error reduction factor. The curves therefore indicate the probability of obtaining a successful herald outcome conditioned on demanding at least a specified degree of indistinguishability distillation.}
    \label{fig:cumulative_plot}
\end{figure}

This motivates considering a minimum required error reduction factor. Figure (\ref{fig:cumulative_plot}) shows the cumulative success probability as a function of this threshold, thereby quantifying the probability of obtaining a successful fusion outcome that also achieves at least a desired degree of indistinguishability improvement.

These results highlight that the conventional notion of fusion success may be incomplete. Depending on the application, successful Bell projections associated with strong anti-distillation behaviour may be undesirable despite satisfying the standard Bell-measurement criterion. This suggests that more granular decision rules in choosing which states to forward to the next step of the computation may be needed.

Interestingly Fig. (\ref{fig:cumulative_plot}) shows that for all conventional fusion protocols (so excluding the altered Grice protocols) the majority of the herald patterns that contribute to the success probability of a Bell-projection are accompanied by unfavourable anti-distillation behaviour under the OBBell error model. While the altered fusion protocols with $\Psi^+$ ancillas have a high success probability when aiming for a minimum error reduction factor of $1$.

Note that Fig. (\ref{fig:cumulative_plot}) reports the possibility to perform a heralded interference experiment that increases the fidelity of Bell states. This generalizes the single-photon distillation protocols and shows that distillation protocols are possible for entangled states too. Such heralded distinguishability distillation of entangled states may be useful in photonic architectures employing deterministic resource-state generation, where entangled photonic states can often exhibit source-dependent distinguishability.

\subsection{Distinguishability propagation in an FBQC-inspired fusion sequence}

To investigate distinguishability propagation in a larger computational setting, we consider a simple model in which the qubits of a 4-GHZ state are sequentially fused with Bell states. This setting captures an essential feature of fusion-based quantum computation, where small entangled resource states are repeatedly connected through large numbers of fusion operations. The 4-GHZ state is particularly relevant since it could act as a resource state for surface-code implementations of FBQC, and in such architectures each qubit of the state ultimately participates in a fusion.

An additional advantage of this setup is that a successful fusion between a 4-GHZ state and a Bell state yields another 4-GHZ state. Consequently, the quality of the resource state after successive fusion operations can be compared directly without the complication that the size or structure of the state changes. This provides a simple setting in which the accumulation of distinguishability through repeated fusion operations can be investigated.

Often, small entangled resource states are created via heralded state generators \cite{bartolucci2021creation, Forbes_2025}. In these cases, an OBB error model as described in Sec. \ref{subsec: Partial dist} is more realistic than the OBBell model as featured in Sec. \ref{sec: OBBell mode, fusing Bell states}. For the results shown in this section, all states, including the ancillary states, are generated from OBB single photons via the GHZ-state generators that use so called primates, described in Ref. \cite{bartolucci2021creation}.

\begin{figure}[h!]
    \centering
    \includegraphics[width=1\linewidth,trim=1.8cm 6.3cm 2.1cm 14.05cm,
    clip]{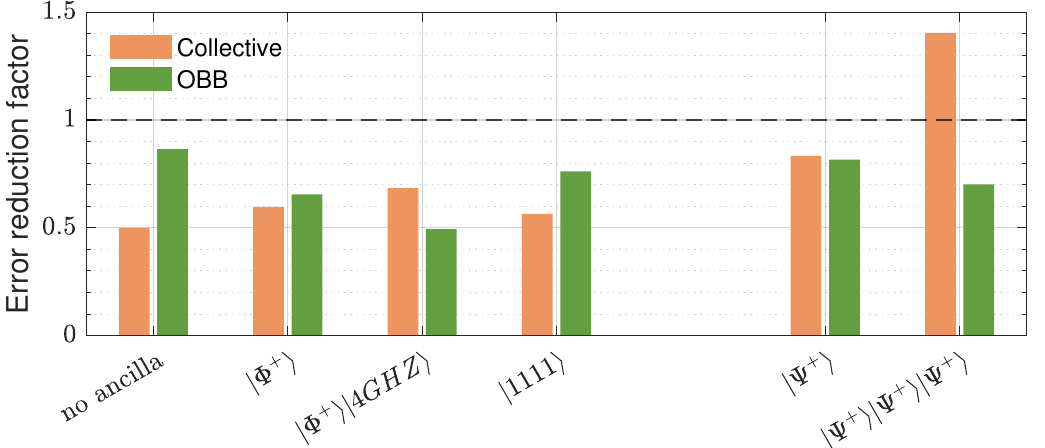}
    \caption{Comparison of the average error reduction factor for the collective distinguishability model (orange) and Bell states generated from OBB single photons (green).}
    \label{fig: comparison histogram}
\end{figure}

Before turning to repeated fusion sequences, it is useful to first isolate the effect of changing the underlying distinguishability model. Fig. (\ref{fig: comparison histogram}) therefore compares the average distinguishability transformation for the fusion of two Bell states under the collective distinguishability model with that obtained when the Bell states (and ancillary states) are instead generated from OBB single photons.

Figure (\ref{fig: comparison histogram}) immediately illustrates that distinguishability transformation strongly depends on the underlying physical error model. Although the altered Grice protocol with $|\Psi^+\rangle^{\bigotimes 3}$ exhibit average indistinguishability distillation under the collective distinguishability model, no comparable improvement is observed when the Bell states are generated from imperfect single photons.

A second important observation is that the qualitative ordering of the fusion protocols changes substantially. In particular, the Ewert \& van Loock protocol performs comparatively poorly under the collective model but becomes one of the most favourable protocols under the OBB model. These observations demonstrate that distinguishability propagation cannot be regarded as an intrinsic property of a fusion protocol alone, but depends strongly on the specific origin of the distinguishability

We next consider repeated fusion operations acting on larger resource states. Following each successful fusion operation, we evaluate the fidelity of the remaining GHZ state. For each protocol, successful herald outcomes are sampled according to their corresponding probabilities. Because the distinguishability transformation depends on the specific sequence of herald outcomes, Monte Carlo sampling is used to capture the stochastic evolution of realistic fusion trajectories. We consider 250 random fusion sequences.

\begin{figure}[h!]
    \centering
    \includegraphics[width=1\linewidth,trim=3.5cm 8.5cm 4.5cm 9cm,
    clip]{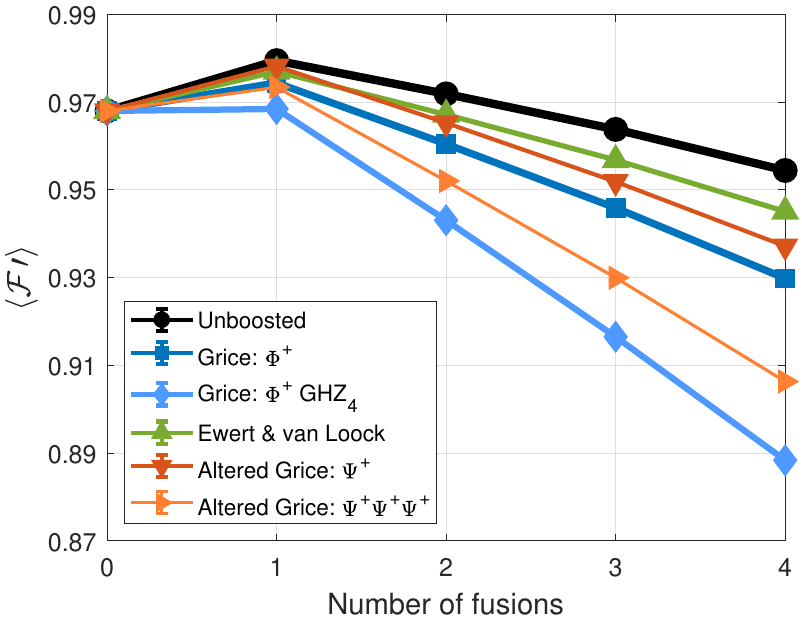}
    \caption{Average fidelity of a 4-GHZ state as a function of the number of successful fusion operations for the various fusion protocols considered. All states (including ancilla states) are generated from OBB single photons. For most data points, the error bars are smaller than the markers. For this figure an input error of $\epsilon_{\mathrm{OBB}}=0.0025$ is used, which follows the reported value of Ref. \cite{Alexander2025}.}
    \label{fig: 4GHZ fused with Bells, average fid}
\end{figure}

Figure (\ref{fig: 4GHZ fused with Bells, average fid}) shows that substantial differences remain between fusion protocols when considering repeated fusion operations on larger resource states. As anticipated from Fig. (\ref{fig: comparison histogram}), however, the qualitative ordering differs from that obtained under the collective distinguishability model. Interestingly, although the altered Grice protocols were identified through optimization under the collective distinguishability model, they continue to exhibit more favourable distinguishability properties than their counterparts employing ancillary states of equal photon number. However, unlike the collective disitnguishability scenario, introducing larger ancillary states now substantially worsens distinguishability propagation. As a result, we find that the unboosted fusion has the most favourable distinguishability propagation. It should be noted that the Ewert \& van Loock protocol exhibits the second best distinguishability propagation in this setting despite not showing advantageous behaviour under the collective distinguishability model.

When looking at Fig. (\ref{fig: 4GHZ fused with Bells, average fid}), we see the same qualitative behavior as Fig. (\ref{fig: comparison histogram}). This gives rise to the hypothesis that the size of the resource state has a smaller impact on distinguishability propagation compared to the error model.

Interestingly, despite the average anti-distillation behaviour shown in Fig. (\ref{fig: comparison histogram}), the fidelity of the surviving 4-GHZ state after the first successful fusion is slightly higher than the fidelity immediately after resource-state generation. This does not contradict the Bell-state analysis. Figure (\ref{fig: comparison histogram}) fuses two Bell states of equal input fidelity, whereas Fig. (\ref{fig: 4GHZ fused with Bells, average fid}) fuses a relatively low-fidelity 4-GHZ state with a considerably higher-fidelity Bell state. For $\epsilon=0.0025$, the initial fidelities are $0.9680$ and $0.9888$ respectively. We therefore attribute the increase in fidelity to the relatively high quality Bell state used in this fusion experiment, outweighing the average anti-distillation associated with the fusion itself.

Most importantly, regardless of the fusion protocol employed, we see a trend in which the fidelity of the resulting quantum state progressively decreases as additional fusion operations are performed. Although different protocols exhibit noticeably different rates of degradation, none of the considered protocols is able to prevent the gradual accumulation of distinguishability.

This observation is particularly relevant for large-scale FBQC, where even a single logical operation requires many successive fusion operations. Consequently, relatively small differences in distinguishability propagation may accumulate. These results therefore suggest that distinguishability transformation properties should be an important consideration when designing fusion protocols for large-scale photonic quantum computation.

\section{Discussion and conclusion}

In this work, we investigated how photonic fusion protocols transform partial distinguishability in heralded entangled states. Rather than characterizing fusion operations solely through Bell-measurement success probability and ancillary resource cost, we studied how distinguishability evolves conditioned on successful fusion events and whether fusion operations can exhibit distillation or anti-distillation behaviour.

Our results show that distinguishability transformation properties constitute an important and largely overlooked aspect of photonic fusion operations. Fusion protocols do not merely determine whether a Bell projection succeeds, but additionally shape the distinguishability properties of the surviving state. Successful fusion events can therefore systematically increase, preserve, or decrease the quality of the remaining state with respect to multiphoton interference.

Importantly, we find that different fusion protocols can exhibit substantially different distinguishability behaviour despite implementing the same Bell-measurement functionality. Consequently, fusion protocols that appear comparable when characterized only by success probability and ancillary resource requirements may nevertheless behave very differently once realistic distinguishability errors are taken into account. In particular, for the conventional fusion protocols considered here, including the standard fusion protocol, the Ewert \& van Loock protocol, and the boosted Grice protocols employing Bell-state and 4-GHZ ancillary states, the average error reduction factor remains below unity under the OBBell error model, indicating net anti-distillation behaviour.

At the same time, we find that modified ancillary-state constructions can significantly alter this behaviour. The altered Grice protocols employing $|\Psi^+\rangle$ ancillary states exhibit considerably more favourable distinguishability properties than their conventional counterparts, despite possessing lower Bell-measurement success probabilities. These observations reveal a trade-off between success probability and distinguishability propagation and demonstrate that ancillary-state design provides an additional degree of freedom in the optimization of fusion protocols.

The importance of these effects becomes increasingly pronounced in large-scale photonic architectures. Fusion-based approaches rely on very large numbers of repeated fusion operations during resource-state generation and computation. Even relatively small systematic biases in distinguishability propagation can therefore accumulate throughout a computation and significantly affect the quality of the surviving entangled states. Consequently, distinguishability transformation properties should be regarded as an important design consideration alongside resource overhead, loss tolerance, and Bell-measurement success probability.

A second important observation is that the qualitative behaviour of fusion protocols depends strongly on the underlying physical error model. Under the collective OBBell model, the different protocols follow a markedly different ordering than in the sequential fusion scenario constructed from OBB photons. This demonstrates that distinguishability propagation depends sensitively on the microscopic origin of the imperfections and cannot generally be inferred from simplified effective error descriptions alone.

More broadly, our results suggest that partial distinguishability should not be viewed as a static source of measurement errors that remains unchanged throughout a computation. Instead, interference and measurements actively transform the distinguishability structure of the photonic state. As a result, the effective error behaviour of a fusion operation depends not only on the distinguishability of the input resources, but also on the specific fusion implementation and the sequence of preceding herald outcomes. In this sense, distinguishability becomes a dynamical property of the computation itself.

This observation may have important implications for effective noise modeling in photonic quantum computation. Several existing approaches model distinguishability-induced imperfections as fixed measurement-error probabilities associated with individual fusion operations. While such descriptions can accurately capture certain measurement statistics, the present results indicate that the underlying distinguishability content of the surviving state evolves throughout the computation due to repeated interference and measurement. Effective measurement-error models based solely on the distinguishability of the initial resource states may therefore fail to capture important correlations and history-dependent effects that emerge during large fusion sequences.

Finally, our work raises several directions for future investigation. Extending the present analysis beyond orthogonal distinguishability models would provide a more refined understanding of distinguishability propagation in realistic devices. It would furthermore be interesting to investigate whether fusion protocols can be explicitly optimized for favourable distinguishability transformation properties while simultaneously retaining high success probabilities. More generally, understanding how realistic imperfections propagate and interact throughout a computational system may contribute to the development of more complete noise models for large-scale photonic quantum architectures.

In conclusion, we have shown that photonic fusion operations possess nontrivial distinguishability transformation properties that significantly affect the quality of heralded entangled states. Our results demonstrate that these properties differ substantially between fusion protocols and depend strongly on the underlying physical error model. Recognizing distinguishability propagation as an additional design consideration therefore provides a more complete framework for the analysis and design of scalable photonic quantum technologies.

\textit{Acknowledgments.}-We thank F.H.B. Somhorst, I.L. Maxwell for scientific discussions. This research is supported by the PhotonDelta National Growth Fund program. This publication is part of the project At the Quantum Edge (VI. Vidi.223.075) of the research programme VIDI which is financed by the Dutch Research Council (NWO).

\textit{Data availability.}-All data used in this study are available in
the 4TU.ResearchData database. \cite{doi.org/10.4121/aff56dab-885b-4b79-872e-f1942a4ffa84}

\bibliographystyle{apsrev}

\bibliography{refs.bib}

\end{document}